\documentclass[aps,pre,twocolumn,10pt,groupedaddress,showpacs,showkeys]{revtex4-1}
\usepackage{graphicx}
\usepackage{amssymb, amsmath}

\begin{document}

\title{Inhomogeneous ensembles of correlated random walkers}

\author{F. Stadler}
\author{C. Metzner} \email{claus.metzner@gmx.net}
\author{J. Steinwachs}
\author{B. Fabry}
\affiliation{Biophysics Group, University of Erlangen, Henkestr.91, D-91052 Erlangen, Germany}

\date{\today}

\begin{abstract}
Discrete time random walks, in which a step of random sign but constant length $\delta x$ is performed after each time interval $\delta t$, are widely used models for stochastic processes. In the case of a correlated random walk, the next step has the same sign as the previous one with a probability $q\!\neq\!\frac{1}{2}$. We extend this model to an inhomogeneous ensemble of random walkers with a given distribution of persistence probabilites $p(q)$ and show that remarkable statistical properties can result from this inhomogenity: Depending on the distribution $p(q)$, we find that the probability density $p(\Delta x, \Delta t)$ for a displacement $\Delta x$ after lagtime $\Delta t$ can have a leptocurtic shape and that mean squared displacements can increase approximately like a fractional powerlaw with $\Delta t$. For the special case of persistence parameters distributed equally in the full range $q\!\in\![0,1]$, the mean squared displacement is derived analytically. The model is further extended by allowing different step lengths $\delta x_j$ for each member $j$ of the ensemble. We show that two ensembles $\left[\delta t, \left\{ (q_j,\delta x_j)\right\}\right]$ and $\left[\delta t^{\prime}, \left\{ (q^{\prime}_j,\delta x^{\prime}_j)\right\}\right]$ defined at different time intervals $\delta t\neq\delta t^{\prime}$ can have the same statistical properties at long lagtimes $\Delta t$, if their parameters are related by a certain scaling transformation. Finally, we argue that similar statistical properties are expected for homogeneous ensembles, in which the parameters $(q_j(t),\delta x_j(t))$ of each individual walker fluctuate temporarily, provided the parameters can be considered constant for time periods $T\gg\Delta t$ longer than the considered lagtime $\Delta t$. Similar models are applicable to many complex systems in which the individual agents undergo distinct - yet aysnchronous - behavioural phases, so that the statistics of the ensemble as a whole can still be considered as stationary. 
\end{abstract}


\keywords{random walks, superstatistics, fluctuation phenomena, noise, random processes}

\maketitle




\section{Introduction}

Many natural phenomena can be described as continuous stochastic processes. The prototypical example for a stochastic phenomenon is Brownian diffusion of a particle within a liquid, usually modeled as a Wiener process. Other systems, such as a particle diffusing within a homogeneous gravitational field, involve a combination of stochastic and deterministic forces. These cases can be modelled by adding a drift term to the stochastic differential equation of the Wiener process. Furthermore, many systems show a tendency to return to a preferred equilibrium state, such as in the case of particle diffusion within a harmonic potential well, which is modeled by the Ornstein$\!-\!$Uhlenbeck process process.

Stochastic processes such as the Wiener or Ornstein$\!-\!$Uhlenbeck process deal with a continuous state variable (e.g., the particle's coordinate $x(t)$) as a function of continuous time. However, when a continuous particle trajectory is measured, the resulting data is typically a time series $x_t$, where the index $t\in\left\{0,1,2,\ldots\right\}$ is counting discrete time points separated by a fixed sample interval $\delta t$. It is most natural to model such time series as a discrete stochastic processes, also called a Discrete Time Random Walk (DTRW) model. While, in general, a DTRW may have a continuous state variable $x\in\Re$, it is often advantageous to further simplify the mathematics and to reduce also the state space of the system to a discrete set, such as $x_t\in\left\{\ldots,-2,-1,0,1,2,\ldots\right\}$. The simplest example is an uncorrelated random walk where the state variable $x_t$ changes after each time interval $\delta t$ by plus or minus a fixed step length $\delta l$. This model can be shown to converge towards the Wiener process in the limit of small $\delta t$ and small $\delta l$.

Possible applications of DTRW models go far beyond traditional physics and include, for example, many biological and economic systems. In particular, the spatio-temporal motion patterns of various biological agents (animals, cells) could be successfully described by DTRW models, but only after correlations between successive steps have been included. In the simplest version of a discrete time correlated random walk (DTCRW), the next step has the same sign as the previous one with a probability $q\!\neq\!\frac{1}{2}$. For values $q\!<\!\frac{1}{2}$, the random walk is called anti-persistent, correponding to a back-and-forth motion. For $q\!=\!\frac{1}{2}$ it behaves like an uncorrelated random walk. For $q\!>\!\frac{1}{2}$, the random walk is called persistent, since it consists of long chains of steps in the same direction. This model is equivalent to a Markov chain and has first been investigated in 1920 \cite{fuerth20}. It is possible to derive analytically the resulting displacement distribution $P(\Delta x, \Delta t)$, which is defined as the probablity that the `particle' has moved a distance of $\Delta x=k\delta x$ during a time interval $\Delta t =n\delta t$, as well as the mean squared displacement $\overline{\Delta x^2}(\Delta t)$ \cite{hanneken98}.

Besides $\delta x$ and $\delta t$, which set the spatial and temporal scale of the system, the only true parameter of this standard correlated random walk model is the persistence parameter $q\!\in\![0,1]$. It directly determines the correlation time $\tau\!=\!\frac{\delta t}{2(1-q)}$ of the velocity autocorrelation function $C_{vv}(\Delta t)\!=\!\sigma_v^2 e^{-\Delta t/\tau}$. This correlation time $\tau$ becomes also apparent in the mean squared displacement $\overline{\Delta x^2}(\Delta t)$ as the characteristic time separating the ballistic regime (small lagtimes) from the diffusive regime (large lagtimes). 

\begin{figure}[!htb]
\includegraphics[width=9cm]{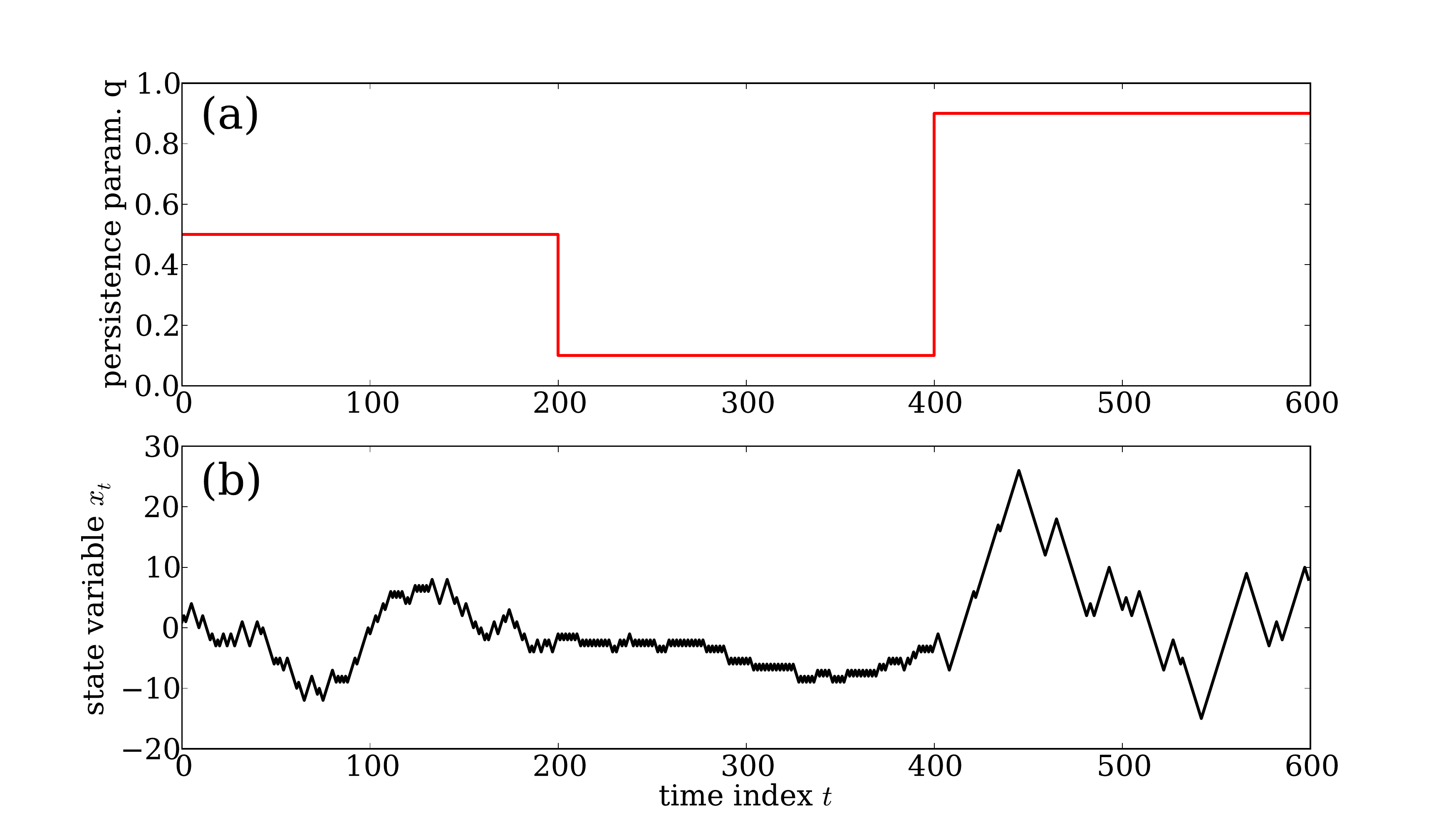}
\caption{\label{fig:phases} Example of a temporally inhomogeneous random walk (b), consisting of three subsequent correlated random walks with different degrees of persistence (a).}
\end{figure}

Not all correlated random walks, however, can be characterized by a single correlation time $\tau$. For example, it is known that individual cells that are migrating on planar substrates can switch between different migration modes, ranging from anti-persistent to highly persistent motion
\cite{potdar10},\cite{niggemann04}. 
The mean squared displacement of such cells often approximates a fractional powerlaw over several decades of lagtime \cite{dieterich08}, pointing to a scale-free random process without any unique correlation time. In order to describe such temporally inhomogeneous systems, it is natural to use so-called superstatistical models \cite{beck03}, in which the statistical parameters (correlation times, diffusion constants, etc.) are themselves subject to random temporal fluctuations (compare Fig.\ref{fig:phases}). 

In this paper, we start from the standard correlated random walk (CRW) model with persistence parameter $q$ and its known quantities $P(\Delta x, \Delta t \;|\;q)$ and $\overline{\Delta x^2}(\Delta t \;|\;q)$. We then generalize the model by allowing $q$ to vary throughout the ensemble, according to a fixed distribution $P(q)$. The resulting displacement distribution $\left\langle P(\Delta x, \Delta t) \right\rangle_q$ and mean squared displacement $\left\langle \overline{\Delta x^2}(\Delta t) \right\rangle_q$ of the ensemble can simply be obtained by computing the average of these q-dependent quantities, weighted with $P(q)$. The average will be performed numerically, revealing almost exponentially shaped displacement distributions and a mean squared displacement resembling a fractional powerlaw over several orders of magnitude in the lagtime. Furthermore, we argue that similar statistical properties are expected for homogeneous ensembles, in which the parameters $(q_j(t),\delta x_j(t))$ of each individual walker fluctuate temporarily, provided the parameters can be considered constant for time periods $T\gg\Delta t$ longer than the considered lagtime $\Delta t$. 

\section{Model and Results}

\subsection{Standard Correlated Random Walk: CRW model}

Since the statistical properties for this model have been derived analytically in Ref.\cite{hanneken98}, we just repeat the main results here. The displacement distribution for a DTCRW with persistence parameter $q$ is given by 
\begin{eqnarray}
&P&(k,n\;|\;q) = P(\Delta x = k\delta x , \Delta t = n\delta t \;|\;q) = \nonumber\\ &=& \sum_{m=1}^{(n-|k|)/2}\! \binom{(n\!+\!k\!-\!2)/2}{m\!-\!1} \binom{(n\!-\!k\!-\!2)/2}{m\!-\!1} \nonumber\\ &\cdot& (1-q)^{2m-1}q^{n-1-2m}  \left( \frac{n(1\!-\!q)+2m(2q\!-\!1)}{2m}  \right),
\end{eqnarray}
where $n$ and $k$ must either both be even or both be odd. It has additionally been assumed that initially (at time step $t\!=\!0$) the probablity for the particle to go left or right are equal. Note that $P(k,n)=0$ for $|k|>n$.

\begin{figure}[!htb]
\includegraphics[width=9cm]{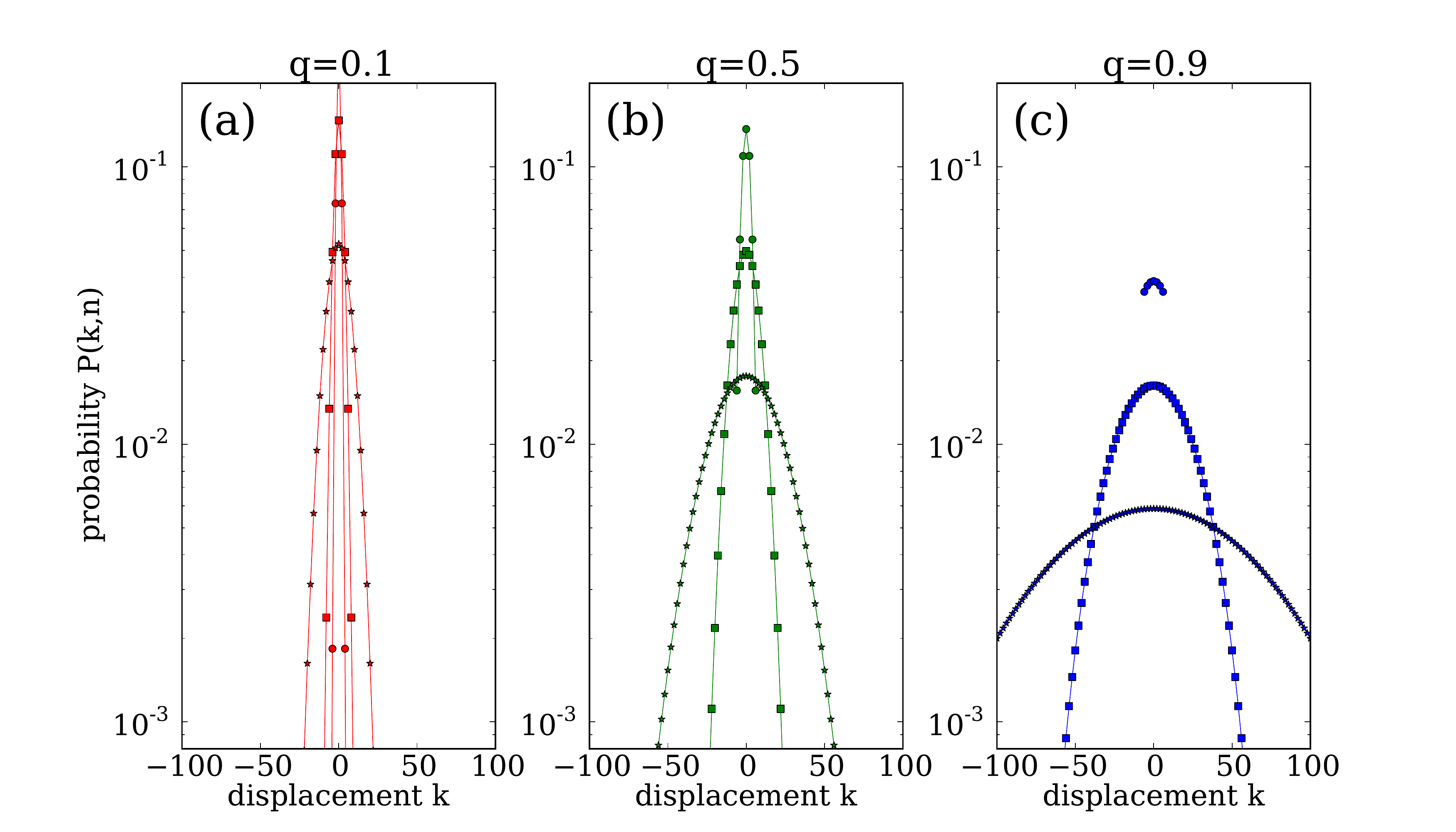}
\caption{\label{fig:swdCRW} Displacement distributions in the correlated random walk model. Case (a) correponds to an anti-persistent walk ($q=0.1$), case (b) to the uncorrelated walk ($q=0.5$) and (c) to a persistent walk ($q=0.9$). In all cases, the distributions are shown for the three lagtimes $n=8$ (circles), $n=64$ (squares), and $n=512$ (stars). The distribution is only defined at the discrete points, lines are guides for the eye. Note that $P(k,n)=0$ for $|k|>n$.}
\end{figure}

Fig.\ref{fig:swdCRW} shows the evolution of the displacement distribution with lagtime for an anti-persistent, the uncorrelated and a persistent case. For lagtimes much larger then the correlation time, the distributions approach Gaussians. The mean squared displacement is given by
\begin{eqnarray}
&&\overline{\Delta x^2}(n\;|\;q) = \overline{\Delta x^2}(\Delta t =n\delta t \;|\;q ) =\nonumber\\
&&\delta x^2 \frac{nq}{1-q}\left\{ 1-\frac{(2q-1)\left[1-(2q-1)^n\right]}{2nq(1-q)} \right\}.
\end{eqnarray}

\begin{figure}[!htb]
\includegraphics[width=7cm]{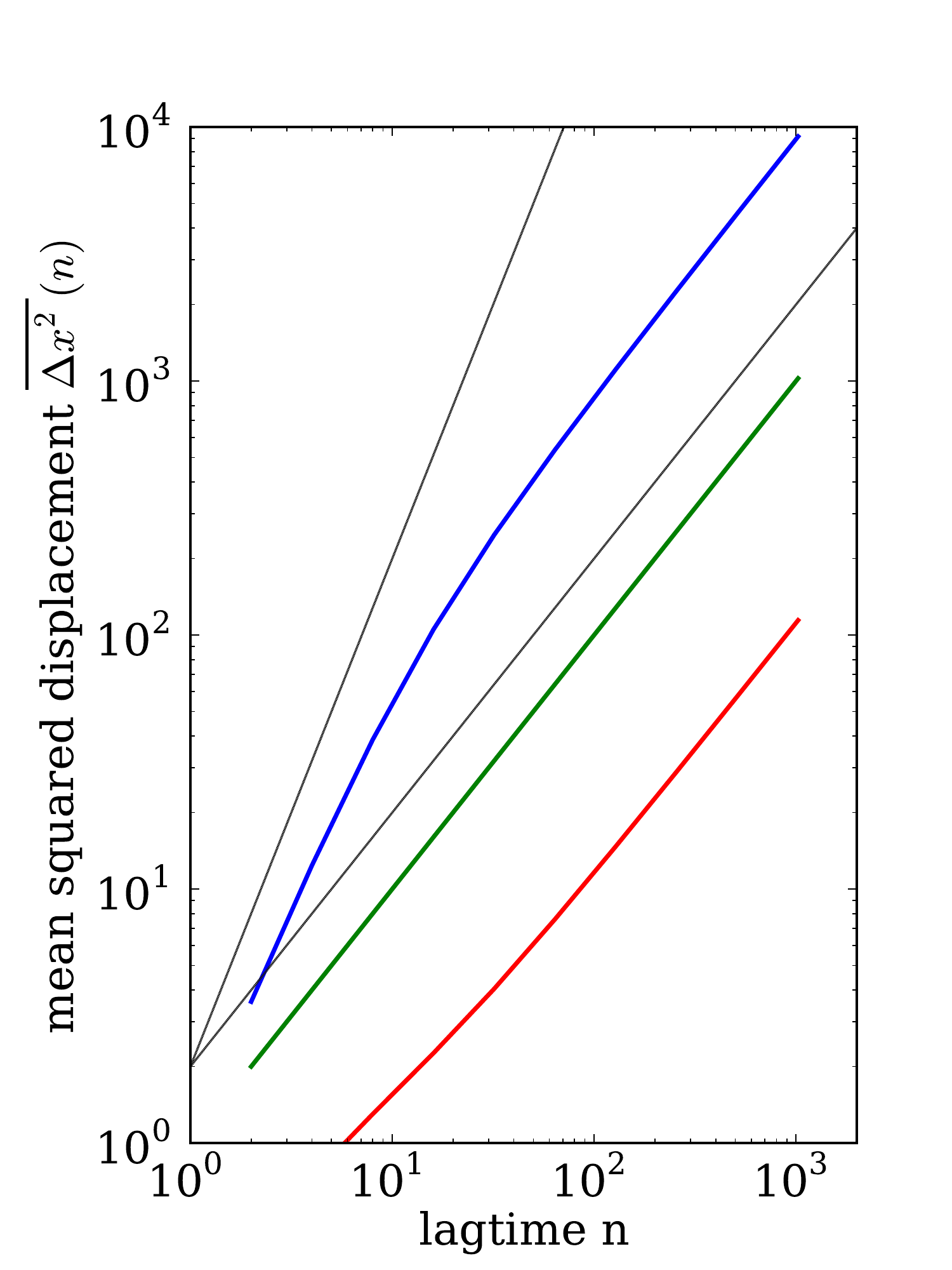}
\caption{\label{fig:msdCRW} Mean squared displacement in the correlated random walk model. The red line correponds to an anti-persistent walk ($q=0.1$), the green line to the uncorrelated walk ($q=0.5$) and the blue line to a persistent walk ($q=0.9$), just as in Fig.\ref{fig:swdCRW}. The thin black lines show a linear and a quadratic dependence.}
\end{figure}

Fig.\ref{fig:msdCRW} shows the mean squared displacement (MSD) as a function of lagtime for the same three degrees of persistence as in Fig.\ref{fig:swdCRW}. In the uncorrelated case (green line), the MSD is linear, corresponding to diffusive random motion. An anti-persistent walk (red line) with the same step length has a smaller MSD, compared to the diffusive case. For a persistent random walk (blue line), the MSD starts ballistically for lagtimes smaller than the correlation time and then continues to grow diffusively.
\vspace{0.5cm}

\subsection{Inhomogeneous Ensemble}

We now assume an inhomogeneous ensemble of correlated random walkers $j$ with different values $q_j$, distributed according to a given distribution $p(q)$. We can then compute the statistical quantities of the ensemble by averaging over the corresponding quantities of the standard correlated walk, so that

\begin{equation}
\left\langle P(k,n) \right\rangle_q = \int_{q\!=\!0}^1 dq \;P(q)\; P(k,n\;|\;q)
\end{equation}

and

\begin{equation}
\left\langle \overline{\Delta x^2}(n\;|\;q) \right\rangle_q = \int_{q\!=\!0}^1 dq \;P(q)\; \overline{\Delta x^2}(n\;|\;q).
\end{equation}

There are specific choices for the distribution $P(q)$ of peristence parameters for which the integrals can be calculated analytically. For example, in the case of an equal distribution $P(q)=1$ in the full range $q\!\in\![0,1]$, the averaged mean squared distribution is given by

\begin{eqnarray}
&&\left\langle \overline{\Delta x^2}(n\;|\;q) \right\rangle_{q,P(q)=1}  =\nonumber\\  
&=& \epsilon + (n+1)\cdot\left[ \Psi\left(\frac{n+3-\epsilon}{2}\right) - \Psi\left(\frac{3}{2}\right) \right].
\label{eq:Analyt}
\end{eqnarray}

Here, $\epsilon\!=\!0$ for even times $n$ and $\epsilon\!=\!1$ for odd times $n$. The digamma function $\Psi(x)=\frac{d}{dx} \log\left(\Gamma(x)\right)$ is defined as the logarithmic derivative of the gamma function. For a derivation of Eq.(\ref{eq:Analyt}) see Appendix.

\begin{figure}[!htb]
\includegraphics[width=9cm]{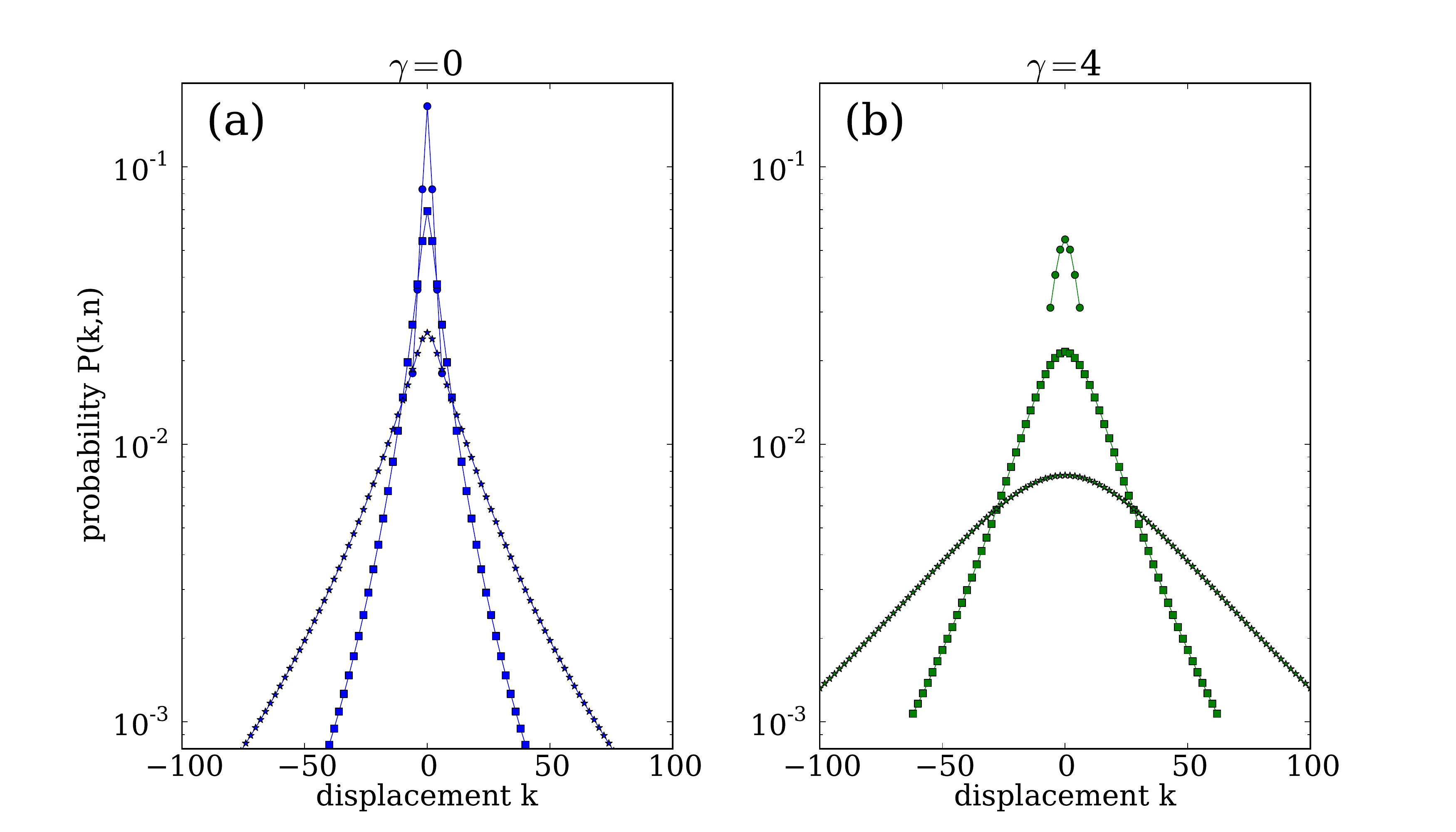}
\caption{\label{fig:swdFPM} Displacement distributions for an inhomogeneous ensemble of correlated random walkers. Case (a) corresponds to equally distributed degrees of persistence with $P(q)=1$, case (b) to a distribution $P(q)\propto\gamma^4$. In all cases, the distributions are shown for the three lagtimes $n=8$ (circles), $n=64$ (squares), and $n=512$ (stars). The distribution is only defined at the discrete points, lines are guides for the eye. Note that $P(k,n)=0$ for $|k|>n$.}
\end{figure}

We have also investigated distribution of the form $P(q)\propto q^{\gamma}$ for the persistence parameter $q\!\in\![0,1]$ and computed the averages numerically for the cases $\gamma=0$ and $\gamma=4$. The resulting distributions (Fig.\ref{fig:swdFPM}) are strongly leptocurtic and resemble exponential functions, rather than Gaussians, which is usually interpreted as a signal of anomaleous random motion. 

\begin{figure}[!htb]
\includegraphics[width=7cm]{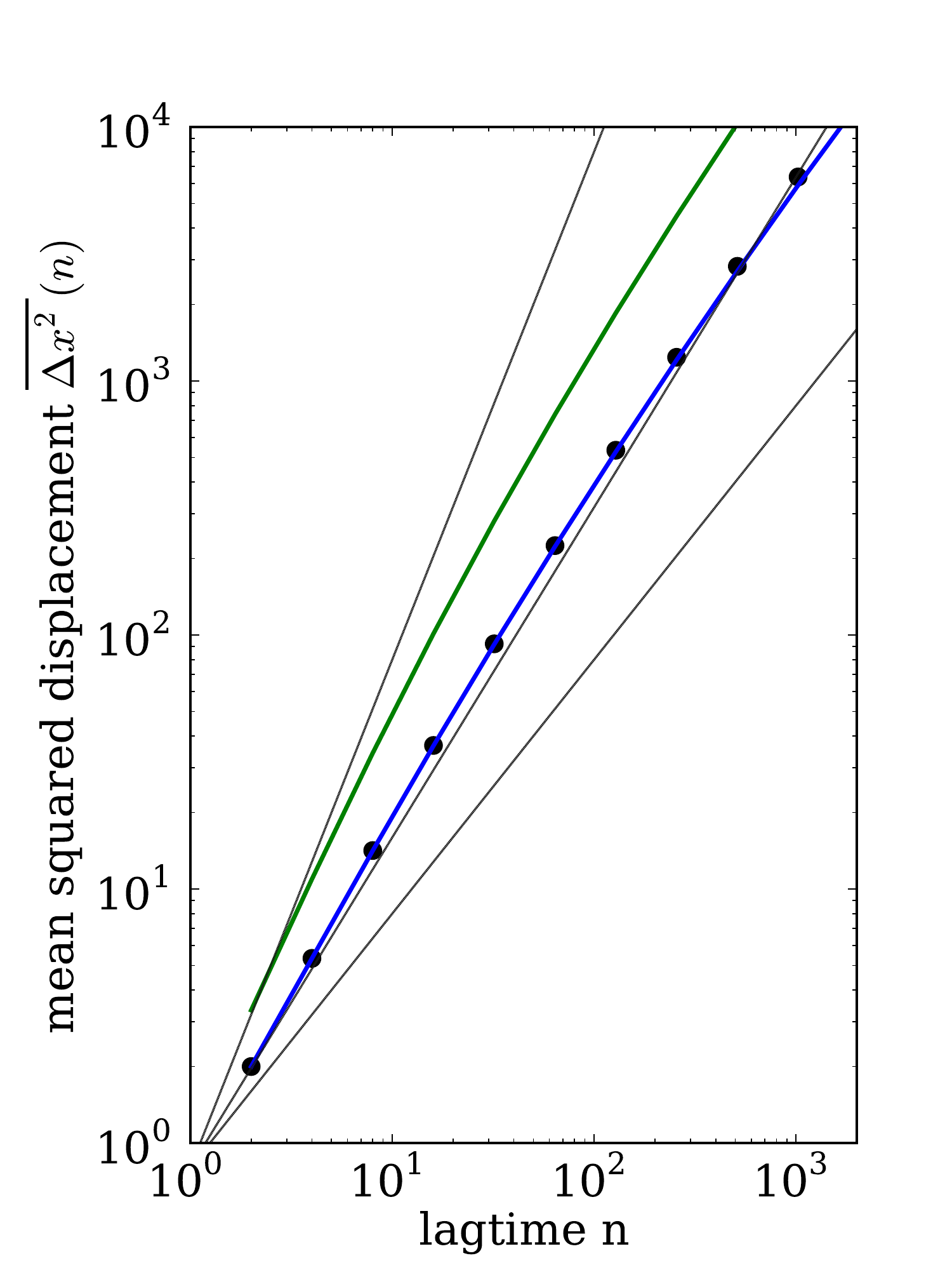}
\caption{\label{fig:msdFPM} Mean squared displacement for an inhomogeneous ensemble of correlated random walkers, corresponding to the distributions $P(q)\propto q^{\gamma}$ of Fig.\ref{fig:swdFPM}. The blue line despicts the case $\gamma=0$, the green line the case $\gamma=4$. The dots show the analytical result according to formula Eq.(\ref{eq:Analyt}). The thin black lines show the powerlaws $n^1$, $n^{1.3}$ and $n^2$.}
\end{figure}

The corresponding mean squared displacements 
(Fig.\ref{fig:msdFPM}) also show clear signatures of anomaleous behavior. In particular, for the case of equally distributed degrees of persistence (blue line), the MSD can be well approximated by a powerlaw with a fractional exponent of about $1.3$ (middle dashed line). In order to confirm our numerical averaging procedure, we have also compared this MSD curve with the analytical result from Eq.(\ref{eq:Analyt}) and found excellent agreement (black dots).

\subsection{Extension of the model and scaling poperties}

A natural extension of the above model is an inhomogeneous ensemble where each walker has not only its own persistence parameter $q_j$, but also an individual step length $\delta x_j$. This extension is actually required for transforming a given ensemble of random walkers from sampling interval $\delta_t$ to another sampling interval $\delta_t^{\prime}$, so that the statistical properties (such as $P(\Delta x, \Delta t \;|\;q)$ and $\overline{\Delta x^2}(\Delta t \;|\;q)$) remain invariant: It has been shown in \cite{met12b} that a single DTCRW with parameters $\left[ \delta t, q, \delta x \right]$ has the same statistical properties as a DTCRW with rescaled parameters $\left[ \delta t^{\prime}=\delta t/s, q^{\prime},\delta x^{\prime} \right]$, if the following scaling relations are used:
\begin{equation} \label{scaleq}
q^{\prime}=q^{\prime}(q,s) = \frac{1}{2}\left( 1\pm \left| 2q-1  \right|^{1/s}  \right), \mbox{and}
\end{equation}
\begin{equation} \label{scalex}
\delta x^{\prime} = \delta x\cdot g(q,s)\;\;\mbox{with}\;\; g(q,s) = \sqrt{\frac{q(1-q)}{q^{\prime}(1-q^{\prime})} \frac{1}{s}  }.
\end{equation}

\begin{figure}[!htb]
\includegraphics[width=7cm]{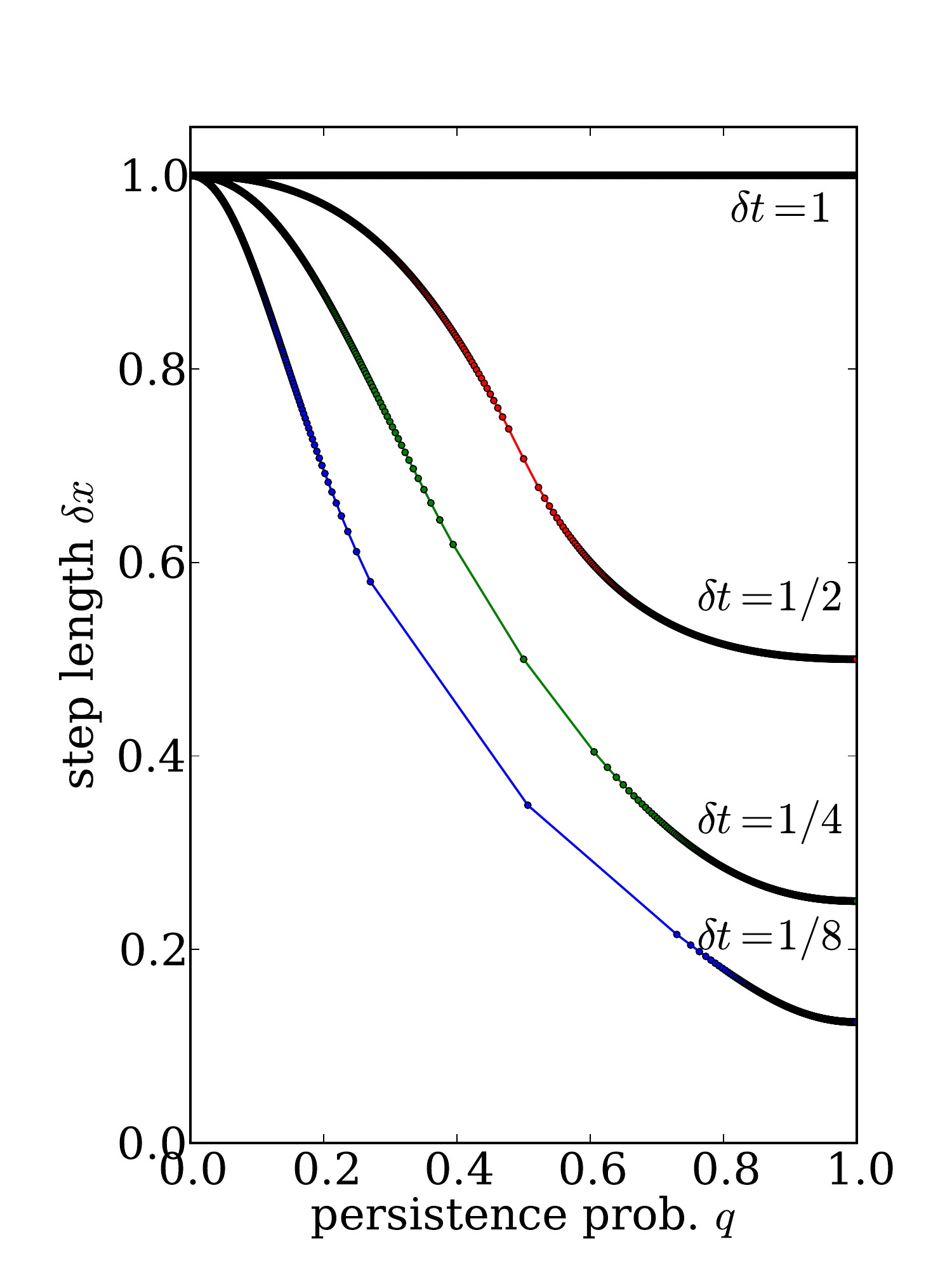}
\caption{\label{fig:ensTrafo} Scaling transformation of an inhomogeneous ensemble of correlated random walkers to different sampling intervals $\delta t$, using the scaling transformations Eq.\ref{scaleq} and Eq.\ref{scalex}. Each individual member $j$ of the ensemble is represented as a point $(q_j,\delta x_j)$. Solid lines are guides for the eye. For $\delta t=1$, the persistence parameters $q_j$ are equally distributed in [0,1] and all step lengths $\delta x_j$ are set to 1.}
\end{figure}

\begin{figure}[!htb]
\includegraphics[width=8cm]{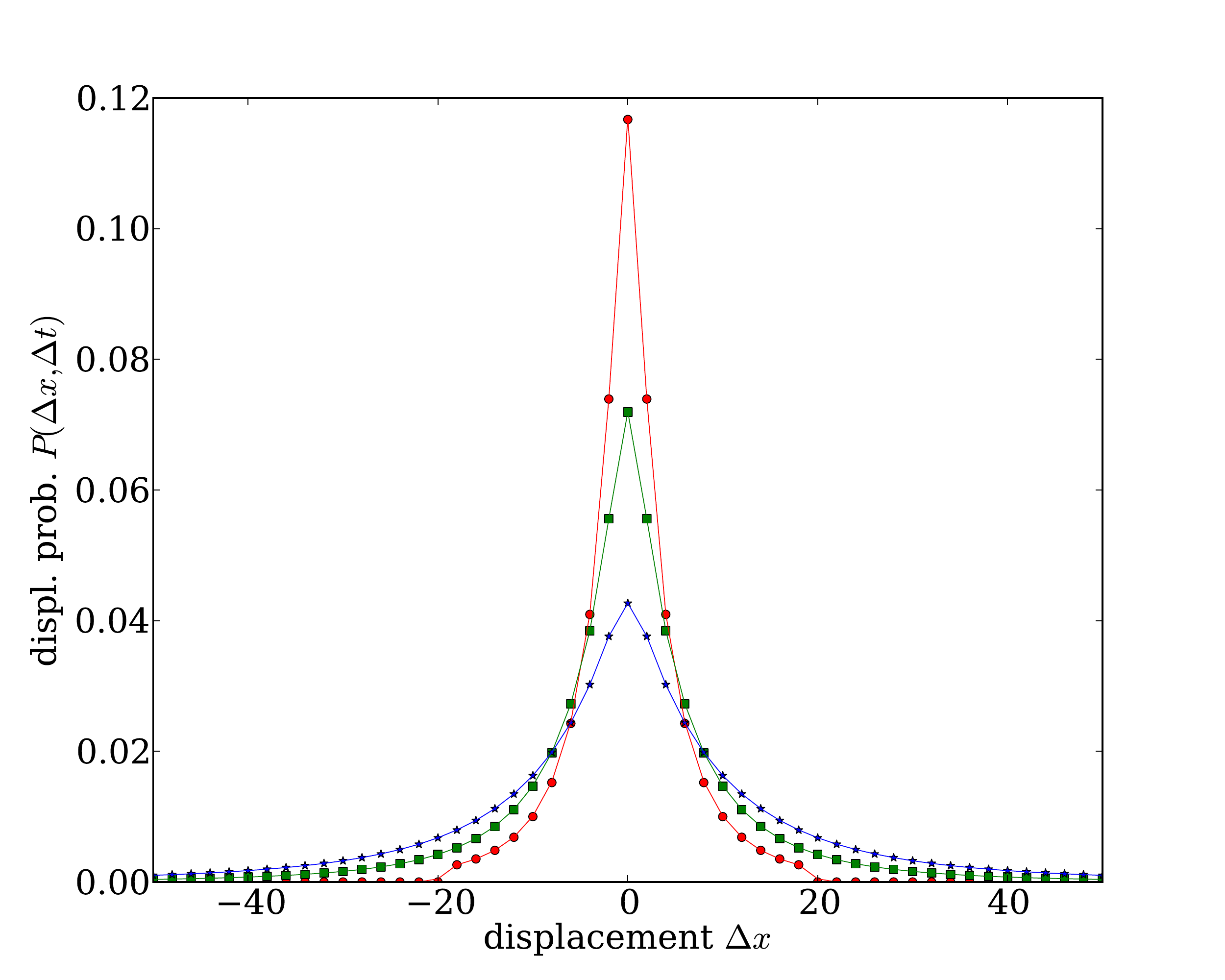}
\caption{\label{fig:swdTrafo} Displacement distributions of an inhomogeneous ensemble of correlated random walkers for three different lagtimes. Solid lines correspond to the ensemble sampled at time increments of $\delta t=1$. The dots show that the distributions remain invariant when the ensemble is rescaled to a smaller sampling time $\delta t=1/8$ according to the scaling transformations Eq.\ref{scaleq} and Eq.\ref{scalex}.}
\end{figure}

These scaling relations have to be applied to each individual member of the ensemble. When transforming the ensemble to a smaller sampling time $\delta t^{\prime}<\delta t$, all step sizes decrease, the anti-persistent walkers become even more anti-persistent and the persistent walkers even more persistent (compare Fig.\ref{fig:ensTrafo}). We have tested the correctness of the scaling transformation by comparing the displacement distributions of an ensemble sampled at $\delta t = 1$ and the corresponding rescaled ensemble sampled at $\delta t = 1/8$ (compare Fig.\ref{fig:swdTrafo}).

\subsection{Ensembles of temporally inhomogeneous walkers}

Above, we have considered an inhomogeneous ensemble of random walkers with temporally constant properties. However, similar statistical properties would be expected in the case of a homogeneous ensemble of random walkers with temporally fluctuating properties. For a concrete example, consider a Petri dish with a set of migrating cells, all of the same type and prepared in the same way. Assume that each specific cell $j$ of this ensemble switches its persistence parameter $q$ in regular intervals $T=N\delta t$ according to the distribution $p(q)$ and produces a time-series $x_t^{(j)}$ corresponding to the x-coordinate of its trajectory. Such an individual time-series $x_t^{(j)}$ does not represent a stationary random process, since the statistical properties (such as the displacement probability $P(x_{t+m}^{(j)}-x_t^{(j)})$) are not time-invariant but depend on the momentary $q$-value of cell $j$. However, if the individual cells switch their $q$-values in a completely independent, asynchronous way, we can compute the ensemble-average ($\left\langle P(x_{t+m}^{(j)}-x_t^{(j)})\right\rangle_j$) and thus obtain a stationary quantity. The $q$-average used above is just a convenient way to compute this ensemble-average. 
 
\section{Discussion and Outlook}

An important point to be addressed is the nature of the approximate `powerlaw' found in specific cases such as shown in Fig.\ref{fig:msdFPM}. It is clear that our random process is not really long-time-correlated and scale-invariant (not even after performing the ensemble-average) since all correlations are lost for lagtimes larger than the duration $T$ of each constant-$q$-phase. Nevertheless, this duration $T$ can last over several decades of lagtime and may exceed the maximum measured time interval in an experiment.

In this paper, we have considered just a very simple case of a superstatistical random walk. The basic concept of starting with a homogeneous random walk model, controlled by a set of parameters $q_1,q_2,\ldots$, and then defining a second layer of random `super-process' for the dynamics of these control parameters, leads to a multitude of possible model variants. These variants may be particularly useful for biological systems with extremely inhomogeneous ensembles. An interesting problem will be how to extract from a measured superstatistical time-series the properties of the super-process from those of the underlying random walk.

\section{Appendix}

We consider a DTCRW model with $\delta x=1$, $\delta t=1$ and persistence parameter $q$. The velocity autocorrelation function is given by
\begin{equation}
C_{\nu\nu}(n\;|\;q) = \frac{ \left\langle \nu_m\;\nu_{m+n}\right\rangle_m } { \left\langle \nu^2_m\right\rangle_m } = (2q-1)^n.
\end{equation}
For the case of equally distributed degrees of persistence, $p(q)=1$, it is straight forward to calculate the average autocorrelation function
\begin{equation}
C_{\nu\nu}(n)=\int_0^1\;(2q-1)^n\; dq = \frac{\frac{1}{2}(1+(-1)^n)}{n+1} = \frac{E_m}{n+1},
\end{equation}
where $E_m=1$ for m even and $E_m=0$ for m odd.

From the autocorrelation function, we can calculate the mean squared displacement by the general relation
\begin{equation}
\overline{\Delta x^2}(n)= \delta x^2\;\sum_{m=-n}^{+n}\;C_{\nu\nu}(m)(n-|m|),
\end{equation}
or
\begin{equation}
\overline{\Delta x^2}(n)/\delta x^2 = n+2n\sum_{m=1}^{n}\;C_{\nu\nu}(m)-2\sum_{m=1}^{n}\;m C_{\nu\nu}(m).
\end{equation}
In our special case, we obtain
\begin{equation}
\overline{\Delta x^2}(n)/\delta x^2 = n+2n\sum_{m=1}^{n}\;\frac{E_m}{m+1}-2\sum_{m=1}^{n}\;m \frac{E_m}{m+1}.
\label{anaMSD}
\end{equation}
The first sum can be rewritten as
\begin{equation}
A_n=\sum_{m=1,2,3,\ldots}^{n}\;\frac{E_m}{m+1} = \sum_{m=2,4,6,\ldots}^{n}\;\frac{1}{m+1}.
\end{equation}
Assuming that the time step $n$ of interest is even, we obtain
\begin{eqnarray}
A_n &=& \sum_{m=1,2,3,\ldots}^{n}\;\frac{E_m}{m+1} = \sum_{k=1,2,3,\ldots}^{n/2}\;\frac{1}{2k+1} \nonumber\\
&=& \frac{1}{2}\left[\Psi(\frac{n+3}{2})-\Psi(\frac{3}{2})\right],
\end{eqnarray}
where $\Psi(x)$ is the digamma function.
The second sum can be rewritten as follows:
\begin{eqnarray}
\sum_{m=1,2,3,\ldots}^{n}\;\frac{m\;E_m}{m+1} &=& \sum_{m=2,4,6,\ldots}^{n}\;\frac{m}{m+1}= \sum_{k=1,2,3,\ldots}^{n/2}\;\frac{2k}{2k+1}\nonumber\\
&=& \frac{n}{2} - \sum_{k=1,2,3,\ldots}^{n/2}\;\frac{1}{2k+1} = \frac{n}{2} - A_n.
\end{eqnarray}
Inserting the sums into Eq.(\ref{anaMSD}), one obtains
\begin{eqnarray}
\overline{\Delta x^2}(n)/\delta x^2 &=& n + 2 n A_n - 2 \left(\frac{n}{2}-A_n\right)\nonumber\\
&=& (n+1) 2 A_n \nonumber\\
&=& (n+1) \left[\Psi(\frac{n+3}{2})-\Psi(\frac{3}{2})\right].
\end{eqnarray}
The case for odd $n$ can be computed in a similar way.

\begin{acknowledgments}
This work was supported by grants from Deutsche Forschungsgemeinschaft.
\end{acknowledgments}

\bibliography{refs}


\end{document}